\documentclass[runningheads]{llncs}
\usepackage{graphicx}
\usepackage{diagbox}
\usepackage[]{babel}
\usepackage{comment}
\usepackage{multirow}
\begin{document}
\title{Studying Robustness of Semantic Segmentation under Domain Shift in cardiac MRI}
\titlerunning{Robustness against Domain Shifts in cMRI}
%
\author{Peter M. Full\thanks{Corresponding author.}\inst{1, 2}\orcidID{0000-0003-4326-8026} \and
Fabian Isensee \inst{1} \and
Paul F. Jäger \inst{1} \and
Klaus Maier-Hein \inst{1}}
\authorrunning{Full et al.}
%
\institute{Division of Medical Image Computing, German Cancer Research Center (DKFZ), Heidelberg, Germany
\email{p.full@dkfz-heidelberg.de}
\and
Medical Faculty Heidelberg, Heidelberg University, Heidelberg, Germany\\
}
\maketitle              
\begin{abstract}
Cardiac magnetic resonance imaging (cMRI) is an integral part of diagnosis in many heart related diseases. Recently, deep neural networks have demonstrated successful automatic segmentation, thus alleviating the burden of time-consuming manual contouring of cardiac structures. Moreover, frameworks such as nnU-Net provide entirely auto- matic model configuration to unseen datasets enabling out-of-the-box application even by non-experts. However, current studies commonly neglect the clinically realistic scenario, in which a trained network is applied to data from a different domain such as deviating scanners or imaging protocols. This potentially leads to unexpected performance drops of deep learning models in real life applications. In this work, we systematically study challenges and opportunities of domain transfer across images from multiple clinical centres and scanner vendors. In order to maintain out-of-the-box usability, we build upon a fixed U-Net architecture configured by the nnU-net framework to investigate various data augmentation techniques and batch normalization layers as an easy-to-customize pipeline component and provide general guidelines on how to improve domain generalizability abilities in existing deep learning methods. Our proposed method ranked first at the \textit{Multi-Centre, Multi-Vendor \& Multi-Disease Cardiac Image Segmentation Challenge (M\&Ms)}.

\end{abstract}

\section{Introduction}
Cardiac MRI (cMRI) is frequently used in clinical workflow to allow for radiation-free, reliable and reproducible diagnostics regarding heart diseases. Intermediate steps, in which contours of the heart's ventricles are manually delineated for at least two time steps of the cardiac cycle, are necessary to diagnose potential heart diseases. These steps are time-consuming and prone to interrater-variability~\cite{Suinesiaputra}. 
To mitigate this problem fully automatic segmentation approaches have been proposed~\cite{Bernard}. However, so far approaches using deep neural networks (DNN) are commonly validated in a setting where the test data is from the same domain as the training data. A drop in performance is often observed when a trained network is applied to data from another domain (i.e. different scanner vendor or acquisition protocol from other hospitals)~\cite{Zech}. The implied requirement for annotated data from all vendors and centres limits the application of DNNs in real-life clinical scenarios. Recent studies propose to alleviate this burden with domain adaptation~\cite{Sandfort} or transfer learning~\cite{Karani}, where knowledge from a source domain is used to estimate a target domain. Limitations of these approaches are the concomitant technical complexity and the fact that unlabeled or partially labeled data from the target domain is required during training. One approach that circumvents these constraints is domain generalization, which aims to learn domain agnostic features given multiple source domains~\cite{Volpi} at training time. In this work, we hypothesize that domain shifts in cardiac segmentation such as between different vendors or centres are less distinct than in typical domain adaptation or domain generalization tasks (e.g. match clip art with photography) and could potentially be bridged with less specialized techniques such as extensive data augmentation. 

nnU-Net is a fully automated segmentation toolkit that allows experts as well as non-experts in the field of medical image analysis to achieve state-of-the-art results in a variety of segmentation tasks without manual intervention~\cite{Isensee}. Here, we aim to enhance nnU-Net's generalization abilities under target domain shifts without degrading its out-of-the-box usability. To this end, we apply extensive data augmentation (DA) and study the interaction with different normalization layers within the nnU-Net framework. The effect of DA on cross-domain performance on medical images~\cite{Zhang_BigAug,Sandfort} and cMRI~\cite{Chen} has been explored before, but these approaches have not been evaluated on publicly accessible datasets, did not consider interdependencies with regularization techniques such as normalization layers, and did not provide results for 2D as well as 3D models. 

Our contributions are the following: 1) We investigate the effect of common DA techniques and batch normalization (BN) on model robustness under domain shifts in cMRI. 2) We investigate whether pooling data from multiple domains during training improves test time performance on individual domains. 3) Our proposed method builds upon the publicly available nnU-Net ensuring out-of-the-box usability to allow for a kick start into domain-invariant fully automatic cardiac segmentation. Our final models will be integrated into the nnU-Net framework and will be freely accessible\footnote{\url{https://doi.org/10.5281/zenodo.4134721}}, as well as our final submission singularity container\footnote{\url{https://doi.org/10.5281/zenodo.4134879}}.

\section{Methods and Experiments}

\subsection{Data and Evaluation Metrics}
The Multi-Centre, Multi-Vendor \& Multi-Disease Cardiac Image Segmentation Challenge ($M\&Ms$) dataset~\cite{Campello} is used for our experiments. The dataset was acquired with four different vendors, namely Siemens, Philips, General Electric, and Canon (in the following as A, B, C, and D, respectively) and at six different centres. The training dataset comprises short axis cMRI images from 175 patients, of which 150 are annotated. For each annotated patient, pixel-wise manual segmentations of the left ventricular myocardium (LVM), the left ventricular blood pool (LV), and the right ventricular blood pool (RV) are given for the end-diastolic (ED) and end-systolic (ES) frames. The 150 annotated samples were acquired with two different vendors (A and B) at three different centres (1, 2, and 3). The test set comprises  200 patients from four different vendors A, B, C, and D and six different centres. Note that vendor C and D are not represented in the training data. Due to the different vendors and centres, domain gaps within the data are expected. While different annotation guidelines at different centres may inject a bias in ground truth segmentation, we assume that this effect is negligible in a highly standardized setting of a challenge. We expect the main difference in data being due to different MR scanner vendors or cardiac diseases being unevenly represented in the different datasets. In the following we will refer to different data domains, when data was acquired with different vendors (e.g. \textit{domain A} when the data was acquired with a scanner from vendor A). For a brief summary of the training and test data see Tab.~\ref{tab:data_set} for a detailed description we refer to~\cite{Campello}. 

All results will be reported on hold-out validation sets using the dice score as target metric. 

\begin{table}
\caption{Training and test dataset description.}\label{tab:data_set}
\begin{center}
\begin{tabular}{ |c|c|c|c|c|c|  } 
\hline
 & Vendor & Centre & \# studies & Annotations & used for training \\
\hline
\multirow{1}{*}{Training} & A & 1 & 75 & yes & yes \\ 
& \multirow{1}{*}{B} & 2 & 50 & yes & yes \\ 
& & 3 & 25 & yes & yes \\
& C & 4 & 25 & no & no \\
\hline
\multirow{1}{*}{Testing} & \multirow{1}{*}{A} & 1 & 21 & no & no \\
& & 6 & 29 & no & no \\
& \multirow{1}{*}{B} & 2 & 24 & no & no \\
& & 3 & 26 & no & no \\
& C & 4 & 50 &  no & no \\
& D & 5 & 50 & no & no \\
\hline
\end{tabular}
\end{center}
\end{table}

\subsection{Data augmentation}
A default nnU-Net version (\textit{default nnU-Net}), with no code changes at all, is compared to a, what we refer to as, \textit{cMRI baseline nnU-Net} ($BL$), which adopts all DA settings, that have shown promising results for the field of cMRI, as proposed by ~\cite{Chen}.
$BL$ comprises image scaling within a range of [0.7; 1.4], random rotation within a range of [$\pm$ 30 degrees], and random flipping horizontally and vertically. Random cropping is automatically applied by nnU-Net when the image size is larger than the patch size. On top of \textit{BL} different DA transformations are added and their effect to the models' performance is measured. For this we consider transformations in image appearance (e.g. brightness, contrast), image shape (e.g. elastic deformations), noise, and orientation (e.g. rotation, random flipping). The specifications of different DA settings can be seen in Tab.~\ref{DA_settings}.

All of the utilized transformations for our experiments are included in the batchgenerators~\cite{batchgenerators} library and do not require manual implementation. DA is performed on-the-fly during training, no additional hard drive storage is needed. 

\begin{table}
\caption{Data augmentation configurations for initial cross domain experiments. The table is thought so complement Fig.~\ref{cross_domain_bp} and give quantitative information about DA methods. Settings for \cite{Chen} are presented as baseline \textit{BL}. We extend DA by \textit{inv. gamma}: inverse gamma , \textit{gaussian}: Gaussian blur and noise, \textit{multi./additive br}.: multiplicative/additive brightness, \textit{all}: all of the aforementioned, \textit{BL enhanced}: ranges of $BL$ enhanced,  \textit{BL enhanced + br}.:  ranges of $BL$ enhanced and brightness enhanced, heavy data augmentation: combination of different DA settings.}\label{DA_settings}
\begin{center}
\scalebox{0.8}{
\begin{tabular}{|l|l|l|l|l|l|l|l|}
\hline
experiment &  rotation & gamma & inv. gamma & Gaussian & add. brightness & multi. brightness & contrast \\ \hline
default nnU-Net  & $\pm 30$ & $(0.7, 1.5)$ & - & yes & - & $\mu = 0$, $\sigma = 0.1$ & no\\
BL &$\pm 30$ & - & - & no & - & - & no \\
BL enhanced & $\pm 60$ & - & - & no & - & - & no\\
BL enhanced + br. &  $\pm 60$ & - & - & no & (0.6, 1.5) & $\mu = 0$, $\sigma = 0.2$ & no\\
BL + all & $\pm 30$ &  (0.7, 1.5) & (0.7, 1.5) & yes &  (0.7, 1.3) & $\mu = 0$, $\sigma = 0.1$ & yes\\

heavy DA & $\pm 180$ &  (0.6, 1.6) & (0.6, 1.6) & yes & (0.7, 1.3) & $\mu = 0$, $\sigma = 0.3$ & yes \\
\hline
\end{tabular}}
\end{center}
\end{table}

\subsection{Model architecture}
Our experimental setup was built with nnU-Net, a dynamic fully automatic segmentation framework for medical images which is based on the widely used U-Net architecture. nnU-Net has shown impressive results in segmentation tasks for different organs and comes with great out-of-the-box usability\footnote[2]{code available at:  \url{https://github.com/MIC-DKFZ/nnUNet}}.

\subsection{Experiments and Results}

In the following, we refer to \textit{intra-domain} performance when a network was trained and evaluated on the same domain (e.g $A(B)\rightarrow A(B)$) and we refer to \textit{cross-domain} performance when a network trained on data from one domain and evaluated on another domain (e.g. $A(B) \rightarrow B(A)$). Further a \textit{mixed domain} scenario is studied, in which data from both vendor A and B is used during training time. 
We investigate the following research questions: \\\\
\textbf{Effect of data augmentation on cross-domain performance} for both \textit{cross-domain} scenarios $A(B) \rightarrow B(A)$. All 75 patients from vendor A(B) were used for training(evaluation). All nnU-Net configurations are fixed through-out these experiments except for DA settings, as described above. This experiment serves as a baseline for the nnU-Net in a typical clinical setting, in which inference is done on a data domain that differs from the training data domain. 

\begin{figure}
\includegraphics[width=\textwidth]{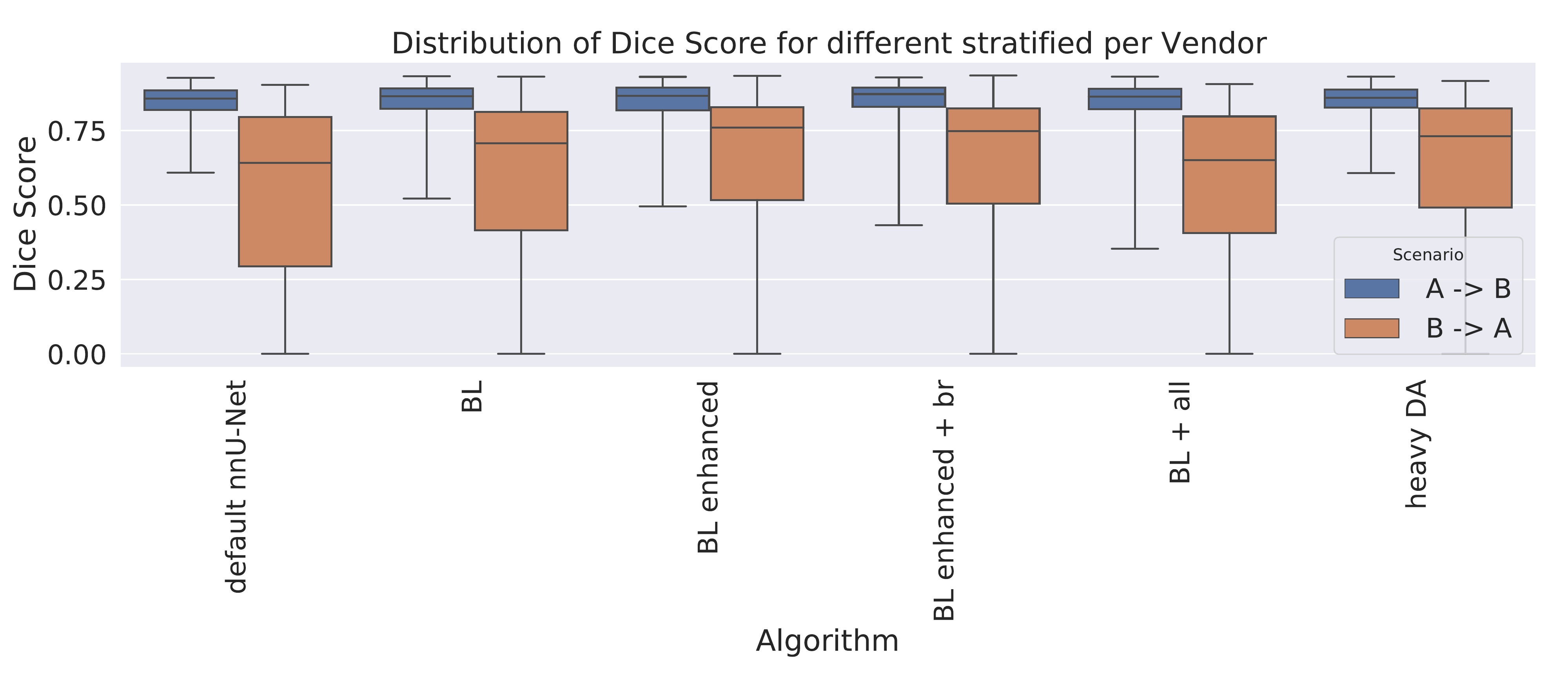}
\caption{\textit{Cross-domain} dice scores for different data augmentation (DA) settings in 2D networks are presented. The blue boxes (left) represent $A \rightarrow B$, with constantly higher dice scores and lower dice score spread for all DA settings compared to $B \rightarrow A$. Note, that for all DA settings some patients achieved a dice score of $0$ for $B \rightarrow A$. A detailed description of DA configuration can be found in Tab.~\ref{DA_settings}}\label{cross_domain_bp}
\end{figure}

Cross-domain results are presented in Fig.~\ref{cross_domain_bp}. While the dice scores for $A \rightarrow B$ are constantly~$>0.8$ (best dice score $0.8548$ for \textit{BL enhanced $+$ br}), $B \rightarrow A$ performs significantly worse for all DA settings (best dice score $0.6622$ for \textit{BL enhanced}). 
This result indicates that DA alone is not able to ensure robust cross-domain performance, especially for $B \rightarrow A$. 

Results from the ACDC challenge suggest that higher dice scores should be achievable~\cite{Bernard}. However, the ACDC challenge reports \textit{intra domain} results. It remains to be investigated whether lower dice scores are to be expected on the M\&Ms data in general or whether the performance gap between ACDC and M\&Ms is due to the differing scenario (\textit{intra} vs \textit{cross domain}). An evaluation on the same validation set is desirable to allow for a fair comparison between models. \\\\
\textbf{Gap between intra- and cross-domain performance}: 25 patients from each vendor are held-out as a validation set. All results will be reported on this fixed evaluation sets. The remaining 50 patients from each vendor are utilized to generate four different training datasets: 1) 50 patients from vendor A, 2) 50 patients from vendor B, 3+4) 25 patients from vendor A and 25 patients from vendor B (\textit{mixed I+II}). With this experimental setup, we study a potential performance gap of $A \rightarrow B$ vs $B \rightarrow A$ on the same hold-out validation set for different DA settings. Furthermore, we investigate whether pooling data from multiple domains during training improves test time performance on individual domains.

The results are shown in Tab.~\ref{tab:four_folds}. Our main observations are that 1) no significant deviations of performance are observed between models trained under the \textit{intra-domain} setting ($A(B) \rightarrow A(B)$) and the \textit{mixed domain} setting ($AB \rightarrow A(B)$). 2) although DA increases the performance for both intra- and cross-domain tasks, the domain gap for $B \rightarrow A$ cannot be closed with our training pipeline. 

Our findings render two possibilities how to improve segmentation results on the target domain: 1) if available add images from the target domain to the training data or 2) if no data from the target domain is available during training, more configurations than just changing DA will have to be optimized. Both will be investigated in the following experiments. \\\\ 
\textbf{Effect of target domain data in the training set} is studied by steadily increasing the number of images from the target domain in the training data. The previous experiment showed that a good segmentation performance was achieved on domain A when a \textit{mixed domain} approach ($AB \rightarrow A$) instead of the \textit{cross domain} approach was used. It remains to be investigated which proportion of images from the initial training set and the target domain are needed to achieve desired segmentation results. Note, in this study we aim to keep the utilized architecture and training procedure fixed throughout all experiments and thus leave more sophisticated approaches such as transfer learning to future research.

\begin{figure}
\includegraphics[width=\textwidth]{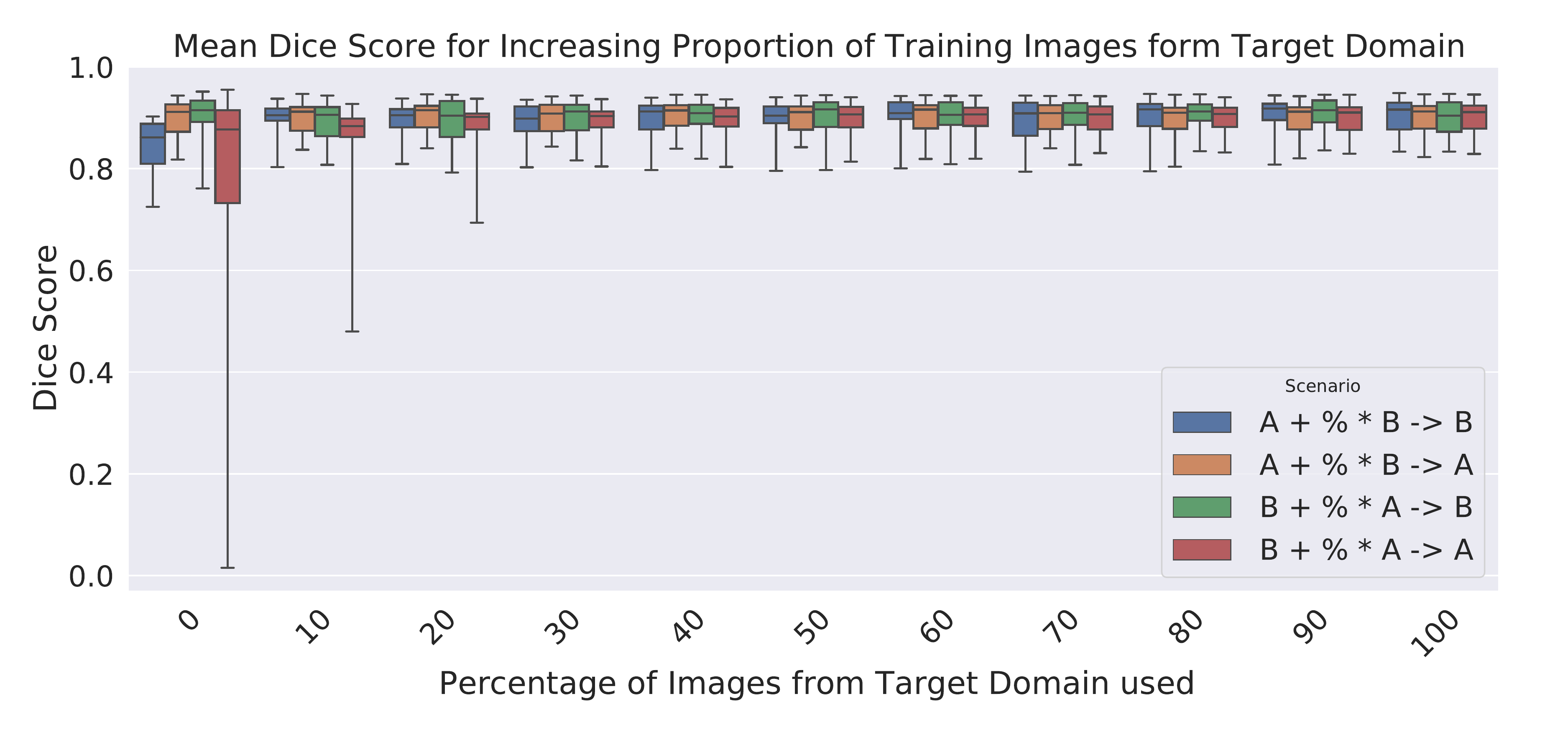}
\caption{Boxplot of mean dice scores for all three classes (LV, LVM, RV). For percentage = 0, only images from the cross-domain setting are included into the training set (100 cross-domain images, 0 images from the target domain). For percentage = 100 all images from the cross-domain setting as well as all images from the target domain are used during training. $B \rightarrow A$ = $B + \% * A \rightarrow A$ for percentage = 0, has been the most problematic scenario in the experiments above. We observe that the mean dice score increases with an increasing number of images from the target domain. For 30\% of images from the target domain, the domain gap between $A \rightarrow B$ and $B \rightarrow A$ is not observable anymore.}\label{Fig:dice_increasing_perc}
\end{figure}

Results are shown in Fig.~\ref{Fig:dice_increasing_perc}. When images from the target domain are added to the training set, cross-domain performance increases until it saturates at $\sim$30\% (15) target domain patients. Beneficial effects of target domain training data are most apparent for $B + \% * A \rightarrow A$, where certain proportions of images from domain A are added to a fixed stack of training images from domain B to predict on images from domain A. Furthermore we observe that the performance for $A + \% * B \rightarrow A$ and $B + \% * A \rightarrow B$ is not compromised even for large proportions of images from a domain that is not the same as the target domain.\\\\
\textbf{Interdependent effects of data augmentation and batch normalization}
were studied in a strict \textit{cross domain} scenario, in which no data from the target was used during training. Instead adding data from the target domain during training, further changes to the nnU-Net configuration were applied. Other studies suggest that introducing BN is beneficial to DNNs generalizability~\cite{Bjorck}. 

To this end, we train a models with little DA (\textit{default nnU-Net}) and extensive DA (\textit{mnms nnU-Net}). Both settings will be trained with 1) the default IN layers of nnU-Net and 2) BN layers. Otherwise the architecture of the network will stay untouched. All results will be reported on a fixed evaluation set of 15 patients per vendor. Our experiments comprise three data scenarios: 1+2) \textit{cross domain} scenarios: Training on 60 patients from vendor A (B) and 3) \textit{mixed domain}: Training on 60 patients from vendor A \textbf{plus} 60 patients from vendor B. All results are provided for 2D networks as well as 3D networks. Configuration details are given in Tab.~\ref{tab:exp_settings}

The results are presented in Tab.~\ref{tab:BN}. Our main observations are: 1) the use of BN layer is especially beneficial, when applied with extensive DA in \textit{cross domain} scenarios (in particular in the previous problematic case $B \rightarrow A$) and 2) \textit{mixed domain} scenarios are on a par with \textit{intra domain} scenarios independent of the chosen normalization technique and DA settings, once DA is extensive enough (compare with Tab.~\ref{tab:four_folds}).

\begin{table}
\caption{Results for networks trained on one domain only (A or B) vs networks trained on images from both domains (mixed I or II), where mixed I/II use different, mutually exclusive subsets from both A and B. The columns hold the mean dice score achieved when evaluated for a specific target domain (A/B) or both domains (A \& B). The four tables represent different data augmentation (DA) setting: upper left: \textit{no DA}; upper right: \textit{BL} DA; lower left and right: two DA settings that go beyond \textit{BL} }\label{tab:four_folds}
\begin{center}
\scalebox{0.85}{
\begin{tabular}{c c}
\begin{tabular}{|l|l|l|l|}
\hline
\diagbox{train on}{test on} &  A & B & A \& B \\ \hline
A &  \textbf{0.6417} & 0.5212 & 0.5815 \\
B & 0.1637 & \textbf{0.6429} & 0.4033\\
mixed I & 0.6115 & 0.6052 & 0.6083 \\
mixed II & 0.6353 & 0.6202 & \textbf{0.6277} \\
\hline
\end{tabular}
&

\begin{tabular}{|l|l|l|l|}
\hline
\diagbox{train on}{test on} &  A & B & A \& B \\ \hline
A &  \textbf{0.8902} & 0.8414 & 0.8658 \\
B & 0.6126 & 0.9022 & 0.7574\\
mixed I & 0.8845 & 0.8996 & 0.8918 \\
mixed II & 0.8857 & \textbf{0.9049} & \textbf{0.8953} \\
\hline
\end{tabular} \\
& \\
\begin{tabular}{|l|l|l|l|}
\hline
\diagbox{train on}{test on} &  A & B & A \& B \\ \hline
A &  \textbf{0.8869} & 0.8412 & 0.8641 \\
B & 0.5879 & \textbf{0.9020} & 0.7449 \\
mixed I & 0.8825 & 0.8940 & 0.8883 \\
mixed II & 0.8800 & 0.8984 & \textbf{0.8892} \\
\hline
\end{tabular}
& 
\begin{tabular}{|l|l|l|l|}
\hline
\diagbox{train on}{test on} &  A & B & A \& B \\ \hline
A &  \textbf{0.8921} & 0.8374 & 0.8647 \\
B & 0.5669 & 0.9044 & 0.7356\\
mixed I & 0.8807 & 0.899 & 0.8899 \\
mixed II & \textbf{0.8921} & \textbf{0.9054} & \textbf{0.8987} \\
\hline
\end{tabular}
\\
\end{tabular}}
\end{center}
\end{table}
\subsection{Final model selection} 
The following observations were considered for the final model submission at the M\&Ms challenge:
1) Tab.~\ref{tab:BN} suggests that doubling training data from 60 to 120 patients (\textit{mixed D}) by combining data from vendor A and B, the segmentation performance is on a par with the \textit{intra D} setting. Therefore we will mix the training data set to achieve good results on the known domains A and B. Assigning images randomly from both vendors, results in a ratio of $\sim$ 1:1, which is well above the threshold of saturation (1:0.3), seen in Fig~\ref{Fig:dice_increasing_perc}. 2) Tab.\ref{tab:BN} suggests that extensive DA combined BN yields large performance improvements especially for the cross domain setting (see \textit{cross D}). Therefore we train our models with BN layers and extensive DA in order to optimize our models for the unseen domains C and D. 

For the final submission we train an ensemble of five 2D and five 3D nnU-Net models using the BN layers together with extensive DA. The trained data is split into five folds comprising 80\% of images from vendor A and B, while the left-out 20\% are mutually exclusive for each fold. 
For 3D networks we will further adjust the resampling configuration. We will set the z-spacing to be the minimal z-spacing from all training cases instead of the nnU-Net default, which (for anisotropic dataset) uses the 10th percentile of the z-spacings found across all training cases.

\begin{table}
\caption{Main differences in data augmentation and normalization layer settings for the baseline \textit{nnU-Net default} and the final submission model \textit{mnms nnU-Net}. Values in brackets represent ranges for the given data augmentation technique. Probabilities that the data augmentation technique will be applied to an image is given as $p=X$. All other architecture features of the nnU-Net were kept unmodified.}\label{tab:exp_settings}

\begin{center}

\begin{tabular}{l|l|l}
Setting & default nnU-Net & mnms nnU-Net \\ \hline
rotation & p=0.2 & p=0.7 \\
elastic deformations & - & p=0.1 \\
independent scale factor per axis & - & p=0.3 \\
elastic deformation alpha & (0, 200) & (0, 300) \\
elastic deformation sigma & (9, 13) & (9, 15) \\
scale & p=0.2 & p=0.3 \\
gamma range & (0.7, 1.5) & (0.5, 1.6) \\
additive brightness  mu & - & 0 \\
additive brightness  sigma & - & 0.2 \\ \hline \hline
normalization layers & IN &  BN \\ 
\end{tabular} 
\end{center}
\end{table}

\begin{table}
\begin{center}
\caption{The effect of batch normalization (BN) compared to instance normalization (IN) for two different DA settings. The results for 2D and 3D are presented. The nnU-net's default DA scheme is presented as \textit{default nnU-Net}. Our model final submission model (\textit{mnms nnU-Net}) showed the best overall performance for 2D and 3D. The three lowest rows in each tabular summarize the six above scenarios into three categories: \textit{intra D} ($A \rightarrow A$ and $B \rightarrow B$), \textit{cross D} ($A \rightarrow B$ and $B \rightarrow A$) and \textit{mixed D} ($AB \rightarrow A$ and $AB \rightarrow B$). For \textit{mixed D} 120 instead of only 60 patients were used for training. Note, that all networks were evaluated on the same hold-out dataset of 15 patients per vendor.}\label{tab:BN}

\begin{tabular}{|l|l|l|l|l|}
\hline
\textbf{2D} & \begin{tabular}[c]{@{}l@{}}default \\  nnU-Net (IN)\end{tabular}  & \begin{tabular}[c]{@{}l@{}}default \\  nnU-Net (BN)\end{tabular} & mnms nnU-Net (IN) & mnms nnU-Net (BN) \\ \hline
$A \rightarrow A$  & 0.8718 & \textbf{0.8832} & 0.8804 & 0.8800  \\
$A \rightarrow B$  & 0.8502 & 0.8594 & 0.8512 & \textbf{0.8735}  \\
$B \rightarrow A$ & 0.6287 & 0.8046 & 0.5809 & \textbf{0.8457}  \\
$B \rightarrow B$ & 0.9121 & \textbf{0.9125} & 0.9063 & 0.905  \\
$AB \rightarrow A$ & \textbf{0.8948} & 0.8911 & 0.8877 & 0.8880  \\
$AB \rightarrow B$ & 0.9066 & \textbf{0.9074} & 0.9069 & 0.9063 \\
& & & &  \\
intra D & 0.8949 & \textbf{0.8983} & 0.8947 & 0.8941  \\
cross D & 0.7394 & 0.8320 & 0.7161 & \textbf{0.8596}  \\
mixed D & \textbf{0.9007} & 0.8993 & 0.8973 & 0.8972  \\
\hline
\end{tabular} 

\begin{tabular}{|l|l|l|l|l|}
\hline
\textbf{3D} & \begin{tabular}[c]{@{}l@{}}default \\  nnU-Net  (IN)\end{tabular} & \begin{tabular}[c]{@{}l@{}}default \\  nnU-Net (BN)\end{tabular} & mnms nnU-Net (IN) & mnms nnU-Net (BN) \\ \hline
$A \rightarrow A$   & 0.8900 & \textbf{0.8914} & 0.8738 & 0.8838  \\
$A \rightarrow B$   & 0.8513 & 0.8640 & 0.8568 & \textbf{0.8761} \\
$B \rightarrow A$  & 0.5668 & 0.6203 & 0.5525 & \textbf{0.7955}  \\
$B \rightarrow B$  & \textbf{0.9112} & 0.9110 & 0.90859 & 0.9098  \\
$AB \rightarrow A$ & 0.8860 & 0.8917 & 0.8832 & \textbf{0.8931}  \\
$AB \rightarrow B$  & 0.9090 & \textbf{0.9104} & 0.90677 & 0.9072  \\
& & & &  \\
intra D & 0.8995 & \textbf{0.9012} & 0.8925 & 0.8979  \\
cross D & 0.7090 & 0.7421 & 0.7046 & \textbf{0.8358}   \\
mixed D & 0.8975 & \textbf{0.9011} & 0.8950 & 0.9001  \\
\hline
\end{tabular} 
\end{center}
\end{table} 

\subsection{Final results on test set}
\begin{figure}
\includegraphics[width=\textwidth]{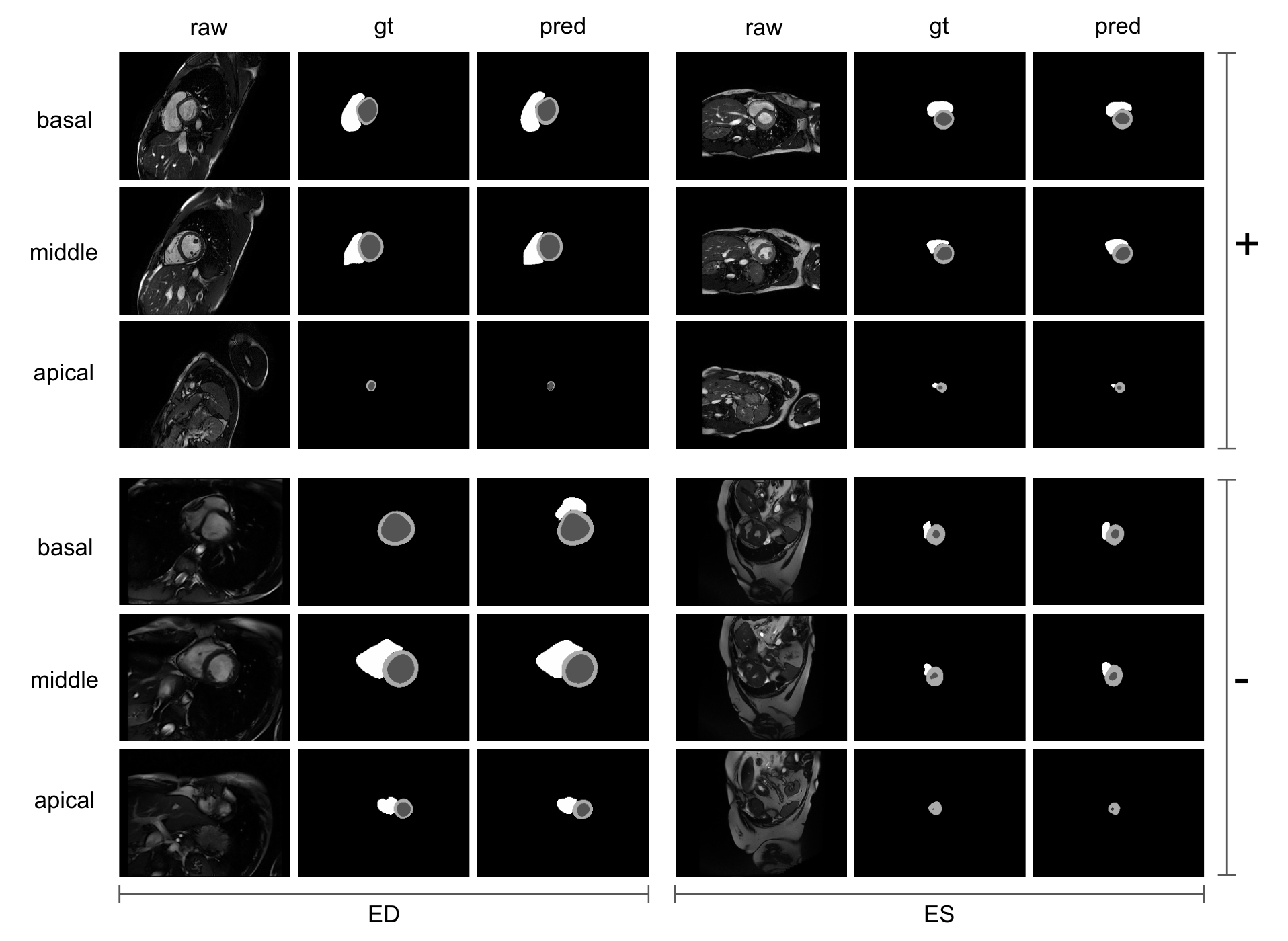}
\caption{Examples of good (+) and poor (-) segmentation results from our final submission for end-diastolic (ED) and end-systolic (ES) time points for basal, mid and apical slices for four different patients from the test set. The definition of \textit{good} and \textit{poor} was given by the challenge organizers. The following vendors are presented: ED(+): vendor D, ES(+): vendor B, ED(-): vendor A, ES(-): vendor C. While the ground truth (gt) and our predictions (pred) are very well aligned for the good cases, some deviations are observed for the poor cases. Comparing the segmentation for ED in row four, we observed that our model predicted a label for the RV, while the human raters did not agree. Note, that automatic and human raters in general disagree the most in the most basal slide, as ambiguities are considered to be most distinct in this part of the heart. Overall good and poor predictions are visually plausible.}\label{fig:examples_test_set}
\end{figure}

The scores for vendor A and vendor B (domains known from training) are slightly above the scores from vendor C and vendor D for LV, LVM, and RV. However we observe that our model achieves a similar score on all four vendors indicating that our submitted model is robust to cMRI data from different vendors and centres in the M\&Ms challenge. Quantitiaive results are shown in Tab.~\ref{final_res}. Qualitative segmentation examples from our test set submission can be seen in Fig.~\ref{fig:examples_test_set}. Note, test set ground truth information was not accessible during the challenge and only made available to us for the purpose of this figure.

\begin{table}
\caption{Mean Dice Score from the final test set stratified for vendors.}\label{final_res}
\begin{center}
\begin{tabular}{|l|l|l|l|l|}
\hline
& Vendor A & Vendor B & Vendor C & Vendor D \\ \hline
LV &  0.9232& 0.9146 & 0.9032 & 0.9091 \\
LVM & 0.8571 & 0.8761 & 0.8418 & 0.8384 \\
RV & 0.8870 & 0.8876 & 0.8838 & 0.8822 \\
\hline
\end{tabular} 
\end{center}
\end{table} 

\section{Conclusion}
We presented an approach to increase robustness against domain shifts in cMRI data from different vendors and centres while at the same time maintaining out-of-the-box usability. This was achieved by 1) building on nnU-Net, a publicly available segmentation framework that achieves state-of-the-art performance on intra-domain segmentation tasks and 2) systematically investigating improve- ments by means of different data augmentation schemes as an easy-to-customize pipeline component and 3) combining BN with extensive DA to achieve significant cross-domain performance gains. 

A limitation of this work is, that we cannot predict whether a data domain will be robustly predicted by a model or not (i.e. why $A \rightarrow B$ works better than $B \rightarrow A$). We will leave this to future research.

We expect these insights to be helpful for typical clinical scenarios in which expert-annotations covering all data domains are not available, but deep learning approaches are to be deployed to minimize cumbersome manual annotation effort.

\section{Acknowledgement}
The authors of this paper declare that the segmentation method they implem- ented for participation in the M\&Ms challenge has not used any pre-trained models nor additional MRI datasets other than those provided by the organizers. Peter M. Full holds a Kaltenbach scholarship from the German Heart Foundation (Deutsche Herzstiftung). Fabian Isensee is funded by the Helmholtz Imaging Platform (HIP).

%
%
%
%

\end{document}